\documentclass[prd, preprint, nofootinbib]{revtex4}
\usepackage{graphicx}
\usepackage{amsfonts}
\usepackage{latexsym}
\usepackage{amssymb}
\usepackage{epsfig}

\begin{document}

\title{ Beginning of Universe through large field hybrid inflation
}

\author{Tatsuo Kobayashi}
 \affiliation{Department of Physics, Hokkaido University, Sapporo 060-0810, Japan
}

\author{Osamu Seto}
 \affiliation{Department of Life Science and Technology,
  Hokkai-Gakuen University, Sapporo 062-8605, Japan
}

%
\begin{abstract}
Recent detection of $B$-mode polarization induced from tensor perturbations 
 by the BICEP2 experiment implies so-called large field inflation, where an inflaton field
 takes super-Planckian expectation value during inflation, at a high energy scale.
We show however, if another inflation follows hybrid inflation,
 the hybrid inflation can generate a large tensor perturbation
 with not super-Plankian but Planckian field value.
This scenario would relax the tension between BICEP2 and Planck
 concerning the tensor-to-scalar ratio, 
 because a negative large running can also be obtained
 for a certain number of e-fold of the hybrid inflation.
A natural interpretation of a large gravitational wave mode
 with or without the scalar spectral running 
 might be multiple inflation in the early Universe.
\end{abstract}

\preprint{EPHOU-14-008}
\preprint{HGU-CAP-032}

\vspace*{3cm}
\maketitle


\section{Introduction}

The detection of $B$-mode polarization
 from gravitational wave mode perturbation 
 has been reported by BICEP2~\cite{Ade:2014xna}.
From the amplitude of the tensor perturbation, the tensor-to-scalar ratio is read as
\begin{equation}
r _T = 0.20^{+0.07}_{-0.05} ,
\end{equation}
 for a lensed-$\Lambda$CDM plus tensor mode cosmological model,
\begin{equation}
r _T = 0.16^{+0.06}_{-0.05} ,
\end{equation}
 after the foreground subtraction based on dust models.
Those values appear to be under the tension with
 the upper bound $r _T < 0.11$ reported by the Planck~\cite{Ade:2013zuv-1,Ade:2013zuv-2}.
As a possible way to resolve this tension, in Ref.~\cite{Ade:2014xna}
 the introduction of a large negative running of the scalar spectral index
 has been proposed.
However, we note that after the BICEP2 paper~\cite{Ade:2014xna}, 
 doubts about inappropriate treatments on dust emissions in their analysis have been raised~\cite{Liu:2014mpa,Flauger:2014qra,Mortonson:2014bja}.
The Planck collaboration have
 released the polarized dust emission data away from Galactic plane~\cite{Adam:2014bub},
 the joint analysis of BICEP2 and the Planck dust polarization data
 found no evidence~\cite{Cheng:2014pxa}.

The large tensor mode has a remarkable implication to inflation.
Such a large tensor mode can be generated in the so-called 
large field inflation models, where 
the field value of inflaton during inflation takes super-Plankian, 
while small field inflation models can not generate large 
$r_T$, but $r_T \leq {\cal O}(10^{-2})$~\cite{LythBound}.
Thus, since the BICEP2 results were announced,
 polynomial chaotic inflation models have
 been studied intensively in light of the BICEP2 data~\cite{ChaoticAfterBicep1,ChaoticAfterBicep2,
ChaoticAfterBicep3,ChaoticAfterBicep4,ChaoticAfterBicep5,ChaoticAfterBicep6,ChaoticAfterBicep7,
ChaoticAfterBicep8,ChaoticAfterBicep9,ChaoticAfterBicep10}.
However, the construction of so-called large field model
 looks nontrivial from supergravity viewpoint as well as field theoretical viewpoint.
For various attempts, see, e.g., Ref.~\cite{Yamaguchi:2011kg-1,Yamaguchi:2011kg-2}. 

In this respect, hybrid inflation is appealing
 since the inflaton $\varphi$ takes a field value
 less than or of the order of Planck scale~\cite{Hybrid-1,Hybrid-2}.
However generally speaking, hybrid inflation models have been disfavored.
Firstly, while simple (non-)supersymmetric hybrid inflation predicts
the density perturbation with the scalar spectral index
 $n_s > (1) \,\, 0.98 $~\cite{Hybrid-1,Hybrid-2,Dvali:1994ms,Linde:1997sj}~\footnote{
In fact, in order to reduce $n_s$, non-minimal Kahler potentials have been
 examined~\cite{NonMinK-1,NonMinK-2}.},
 the WMAP~\cite{Hinshaw:2012aka} and Planck~\cite{Ade:2013zuv-1,Ade:2013zuv-2}
 data indicate $n_s \simeq 0.96$. 
Secondly, topological defects, usually cosmic strings, are formed at the end of a hybrid inflation,
 when the water fall field develops the vacuum expectation value (vev). 
The expected mass per unit length of the formed cosmic string is not compatible
 with data of the temperature anisotropy $\delta T/T$
 in the cosmic microwave background radiation (CMB)~\cite{StringProblem}.
Thirdly, hybrid inflation cannot generate gravitational wave mode
 with a large amplitude~\cite{Civiletti:2014bca-1,Civiletti:2014bca-2,Civiletti:2014bca-3}
 because the potential energy scale is too low, in other word the potential is too flat 
 when we normalize the amplitude of the density (scalar) perturbation with $\delta T/T$.

We point out a possible way to overcome those problems of hybrid inflation.
The second problem due to cosmic strings is avoided
 by considering somewhat complicated potential such as
 shifted hybrid inflation~\cite{Jeannerot:2000sv}
 or smooth hybrid inflation~\cite{Lazarides:1995vr}.
The other two can be simultaneously solved if we give up
 a large enough number of e-folds $N$ by a hybrid inflation
 to solve problems in the standard Big Bang cosmology.
A smaller $N$ corresponds to a larger slow roll parameters, 
 which leads to a smaller $n_s$ and a larger $r_T$.
Of course, we need to solve the horizon and flatness problems
 and generate the density perturbation at super-horizon scales.
This may be achieved by an inflation following after the hybrid inflation,
 so-called double inflation scenario, where the Universe has undergone
 an inflationary expansion in the early Universe more than once.
Such double inflation scenarios~\cite{DoubleInflation} have been considered
 with various motivations, e.g.,
 low multipole anomaly in the CMB sky~\cite{Lowpole-1,Lowpole-2,Lowpole-3},
 the generation of primordial black holes~\cite{PBH1,PBH2-1,PBH2-2,PBH2-3},
 and the dilution of unwanted relics~\cite{Lyth:1995ka-1,Lyth:1995ka-2,Lyth:1995ka-3}.

In this paper, we show if the secondary inflation takes place
 with a sufficient number of e-folds, say $N \simeq 40 - 50$, 
 a kind of supersymmetric hybrid inflation is available and could generate
 not only an appropriate density perturbation and its spectrum
 as in Ref.~\cite{Lazarides:2007dg}
 but also large $r_T = {\cal O}(0.1)$,
 in contrast with the previous work~\cite{Civiletti:2011qg}
 where the possibility of $r_T\simeq 0.02$ has been pointed out.

\section{Double Inflation scenario}

Here, at first, we note several formulas used in the following analysis.
The power spectrum of the density perturbation, 
 the scalar spectral index, its running and the tensor-to-scalar ratio are expressed as
\begin{eqnarray}
{\cal P}_{\zeta} &=& \left(\frac{H^2}{2\pi |\dot{\varphi}|}\right)^2
 = \frac{V}{24 \pi^2 \epsilon}, \\
n_s &=& 1 + 2 \eta -6 \epsilon ,\label{Formula:ns}\\
\alpha_s &=& 16 \epsilon\eta -24 \epsilon^2 -2\xi ,\label{Formula:alphas}\\
r_T &=& 16 \epsilon ,\label{Formula:rT}
\end{eqnarray}
 respectively by using the potential $V$ and slow roll parameters
\begin{eqnarray}
 \eta &=& \frac{V_{\varphi\varphi}}{V} , \\
 \epsilon &=& \frac{1}{2}\left(\frac{V_{\varphi}}{V}\right)^2 , \\
 \xi &=& \frac{V_{\varphi}V_{\varphi\varphi\varphi}}{V^2} ,
\end{eqnarray}
 in the unit of $8 \pi G =1$.
Here, $\varphi$ is the canonically normalized inflaton field, 
 a subscript $\varphi$ and dot denote derivatives with respect
 to $\varphi$ and time respectively,
 and $H$ is the Hubble parameter.

\subsection{Shifted hybrid inflation as the first inflation}

The motivation of the supersymmetric (F-term) hybrid inflation is to 
realize the inflation model through gauge symmetry breaking 
in supersymmetric grand unified theories, e.g. with gauge groups $G$ such as 
$SU(5)$, $SO(10)$ and $SU(4) \times SU(2) \times SU(2)$, 
although here we do not concentrate on a specific gauge group $G$.
The superpotential for the simplest supersymmetric hybrid inflation~\cite{Dvali:1994ms}
 is given by
\begin{equation}
W = \kappa S(\bar{\Phi}\Phi- M^2) ,
\label{Hybrid:Superpotential}
\end{equation}
 with $\kappa$ being a Yukawa coupling.
Here, $S$ is a gauge singlet superfield and 
 $\Phi$ and $\bar \Phi$ are charged superfields, which have representations 
conjugate to each other under $G$.  
When $\Phi$ and $\bar \Phi$ develop their vevs, the gauge symmetry is 
broken.
Although the above superpotential is simple and inflation can be realized, 
this simple hybrid inflation suffers from the stringent constraint on
 the tension of cosmic strings generated at the end of inflation.
Shifted hybrid inflation~\cite{Jeannerot:2000sv} is an attractive alternative
 which is free from cosmic string problem,
 because the symmetry of the water fall field is already broken during inflation.
Nevertheless the field dynamics is same as in the minimal hybrid inflation.
Thus, in the following analysis, we accept shifted hybrid inflation model.

The superpotential for shifted inflation is given by
\begin{equation}
W = \kappa S(\bar{\Phi}\Phi-\mu^2)-S\frac{(\bar{\Phi}\Phi)^2}{M^2} .
\label{Shifted:Superpotential}
\end{equation}
The scalar potential is written as 
\begin{equation}
V = \left|\kappa (\bar{\Phi}\Phi-\mu^2)-\frac{(\bar{\Phi}\Phi)^2}{M^2}\right|^2
 + (|\Phi|^2+|\bar{\Phi}|^2)|S|^2
 \left|\kappa -2\frac{\bar{\Phi}\Phi}{M^2}\right|^2 + V_D,
\label{Shifted:GlobalPotential}
\end{equation}
where $V_D$ denotes the $D$-term potential of $G$.
The scalar potential $V$ is rewritten as
\begin{equation}
V = \left(\kappa (\phi^2-\mu^2)-\frac{\phi^4}{M^2}\right)^2
 + 2 \phi^2|S|^2 \left(\kappa -2\frac{\phi^2}{M^2}\right)^2 ,
\label{Shifted:SimplifiedGlobalPotential}
\end{equation}
 by imposing the $D$-flat condition $\Phi = \bar{\Phi}^* \equiv \phi$.
%
\begin{figure}[!t]
\begin{center}
\epsfig{file=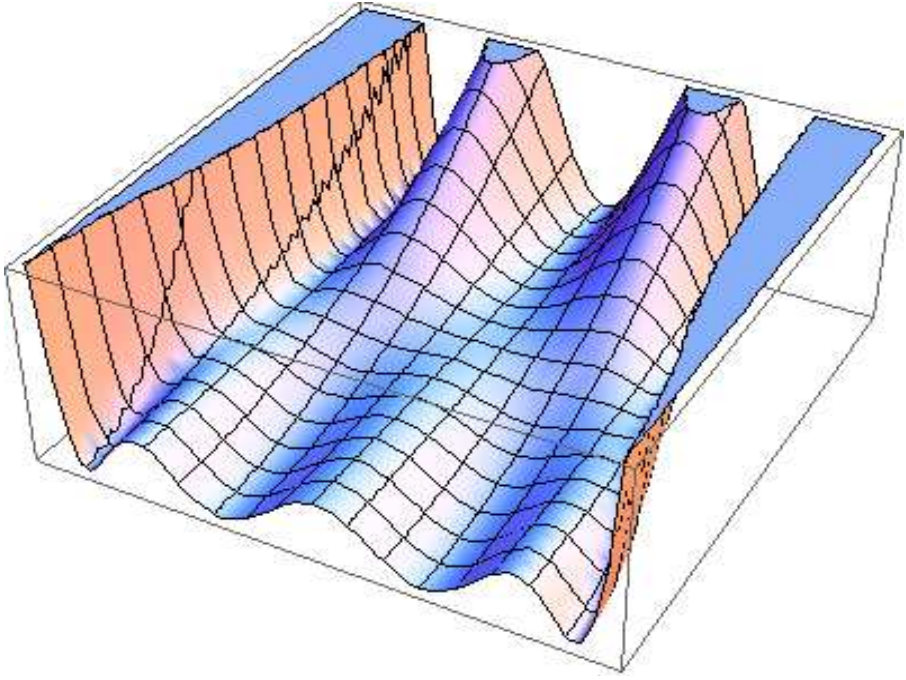, width=12cm,height=9cm,angle=0}
\end{center}
\caption{ The shape of the scalar potential (\ref{Shifted:SimplifiedGlobalPotential}).
 }
\label{Fig:potential}
\end{figure}
%
%
The supersymmetric global minimum is located at
\begin{equation}
 (S, \phi^2)  =  
 \left( 0, \frac{\kappa M^2}{2}\left(1 \pm \sqrt{1-\frac{4\mu^2}{\kappa M^2}}\right) \right),
\end{equation}
if the following condition 
\begin{equation}
 \mu^2 < \frac{\kappa M^2}{4},
\label{Shifted:FirstCondition}
\end{equation}
is satisfied. In this vacuum, 
 the masses of the inflaton $S$ and the second field $\phi$ are given by
\begin{eqnarray}
 m_S^2 = m_{\phi}^2 
 = \kappa M^2\left(1-\sqrt{\frac{\kappa M^2-4\mu^2}{\kappa M^2}}\right).
\end{eqnarray}
Stationary points of this potential with respect to $\phi$ are given by 
\begin{eqnarray}
\phi^2 =0, \frac{\kappa M^2}{2}, \phi^2_{\pm} ,
\end{eqnarray}
where $\phi^2_{\pm}$ is given by
\begin{eqnarray}
\phi^2_{\pm} = 
\frac{\kappa M^2-6|S|^2 
\pm \sqrt{(\kappa M^2-6|S|^2)^2-4\kappa(\mu^2-|S|^2)M^2}}{2}.
\label{def:phipm}
\end{eqnarray}
The squared root term in the right hand side of Eq.~(\ref{def:phipm}) can be real
 only for
\begin{equation}
 |S|^2 > \frac{1}{18}(2\kappa M^2+\sqrt{36\kappa\mu^2 M^2-5\kappa^2M^4}) ,
\end{equation}
or
\begin{equation}
 |S|^2 < \frac{1}{18}(2\kappa M^2-\sqrt{36\kappa\mu^2 M^2-5\kappa^2M^4}) ,
\end{equation}
 if 
\begin{equation}
 \frac{9}{5}\mu^2 > \frac{\kappa M^2}{4} ,
\label{Shifted:SecondCondition}
\end{equation}
 is satisfied.

As can be seen in Fig.~\ref{Fig:potential},
for large $|S|$ region, the potential has three local minima at 
 $\phi = -\sqrt{\frac{\kappa M^2}{2}}, 0, \sqrt{\frac{\kappa M^2}{2}}$
 separated by two local maxima at $-\phi_+$ and $\phi_+$.
We consider the case that
 the inflaton slow rolls in the inflation valley at $\phi^2 = \kappa M^2/2$
 during inflation in order not to produce topological defects.
As $|S|$ decreases, the local maximum $\phi_+$ approaches the valley
 and coincides with it at
\begin{equation}
 |S|^2 = |S_c|^2 \equiv \frac{1}{2}\left(\frac{\kappa M^2}{4} - \mu^2\right) .
\end{equation}
Then, $\phi$ starts to fall into the $\phi_-$ minimum.

Due to supersymmetry breaking during inflation,
 radiative corrections to the scalar potential 
\begin{equation}
\delta V = \frac{\kappa^4\sigma_c^4}{8\pi^2} \ln\frac{\sigma}{\Lambda} ,
\end{equation}
 appears for $S \gg S_c$ in terms of the canonical inflaton $\sigma = \sqrt{2}|S|$.
In addition, supergravity effects also lift the potential.
In total, we consider the scalar potential
\begin{eqnarray}
V = \kappa^2 \sigma_c^4
 \left[1 + \frac{\kappa^2}{8\pi^2} \ln\frac{\sigma}{\Lambda}
 + \frac{1}{2}m^2\sigma^2 + ...\right],
\end{eqnarray}
where the third term  in the right-hand side is a dominant supergravity effect. 
The ellipsis represents higher order supergravity corrections, but in the following analysis
 we assume those terms are not significant for inflationary dynamics.
Note that this is the same form as that of the standard supersymmetric hybrid inflation.
The slow roll parameters are expressed as
\begin{eqnarray}
 \eta &=& - \frac{\kappa^2}{8\pi^2\sigma^2}+m^2, \label{eta:sigma} \\
 \epsilon &=& \frac{1}{2}\left(\frac{\kappa^2}{8\pi^2\sigma}+m^2 \sigma \right)^2 , \label{epsilon:sigma} \\
 \xi &=& \left(\frac{\kappa^2}{8\pi^2\sigma}+m^2\sigma \right)\left(\frac{2 \kappa^2}{8\pi^2\sigma^3}\right) . \label{xi:sigma}
\end{eqnarray}
We substitute Eqs.~(\ref{eta:sigma}) - (\ref{xi:sigma})
 and the solution of $\sigma$ during inflation
\begin{equation}
\int_{\sigma_e}^{\sigma} \frac{V}{V_{\sigma}}d\sigma = N ,
\label{Shifted:solution}
\end{equation}
with $N$ being the number of e-folds and
\begin{equation}
\sigma_e^2 \simeq \frac{\kappa^2}{8\pi^2 (1+m^2)} ,
\end{equation}
 being the final field value
 where the slow roll parameter $\eta$ becomes $-1$ and inflation terminates, 
 for Eqs.~(\ref{Formula:ns}) - (\ref{Formula:rT}).
Then,  we obtain values of slow roll parameters.
The amplitude of the density perturbation provides the normalization of 
 the potential height $\kappa^2 \sigma_c^4$. 

\begin{figure}[!t]
\begin{center}
\epsfig{file=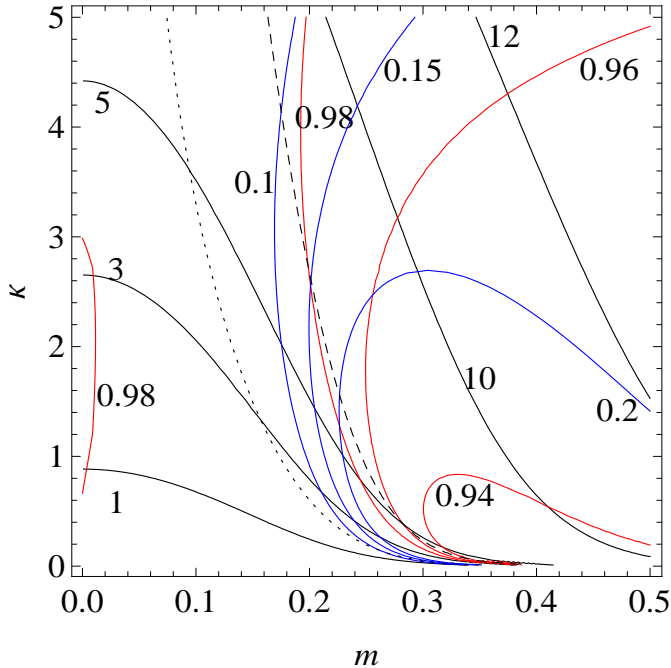, width=9cm,height=9cm,angle=0}
\end{center}
\caption{ Contours for $N=50$; 
Blue lines are for the tensor-to-scalar ratio $r_T$,
and red lines are the spectral index $n_s$. 
For small $\kappa$ and vanishing $m$, 
 the well know prediction of $n_s \simeq 0.98$ and very small $r_T$
 is recovered.
For other contours, see the text.
 }
\label{Fig:50}
\end{figure}

\begin{figure}[h,b]
\begin{center}
\epsfig{file=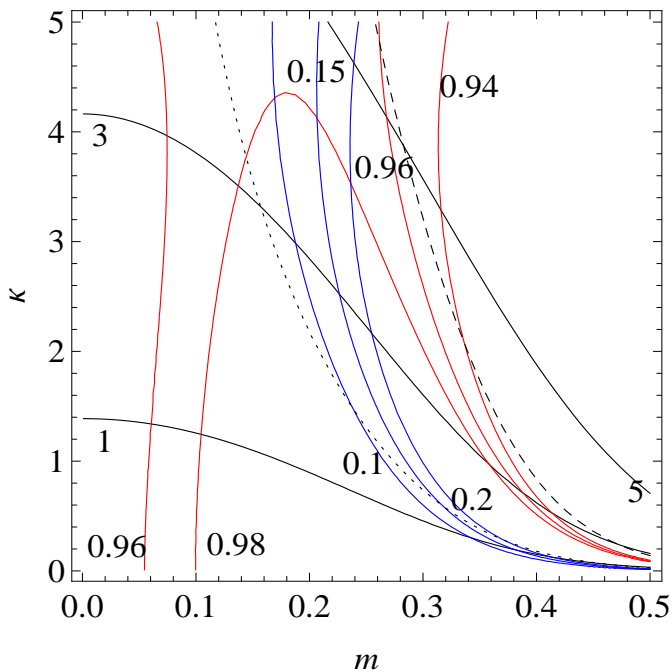, width=9cm,height=9cm,angle=0}
\end{center}
\caption{
Same as Fig.~\ref{Fig:50} but for $N=20$. 
 }
\label{Fig:20}
\end{figure}

\begin{figure}[!t]
\begin{center}
\epsfig{file=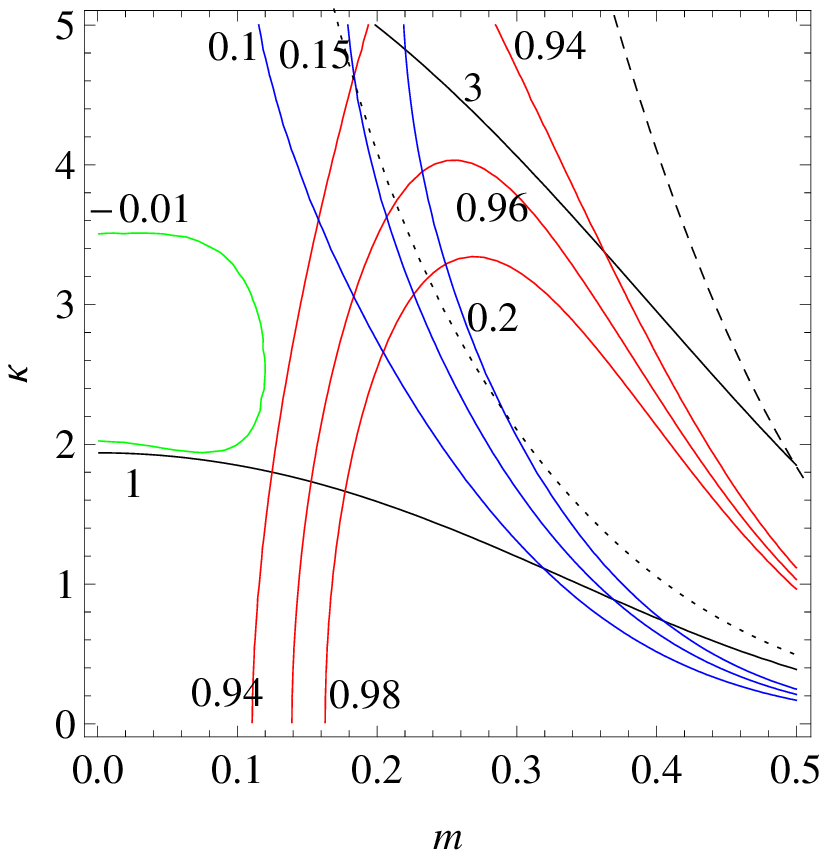, width=9cm,height=9cm,angle=0}
\end{center}
\caption{ Various contours for $N=10$.
 }
\label{Fig:10}
%
\begin{center}
\epsfig{file=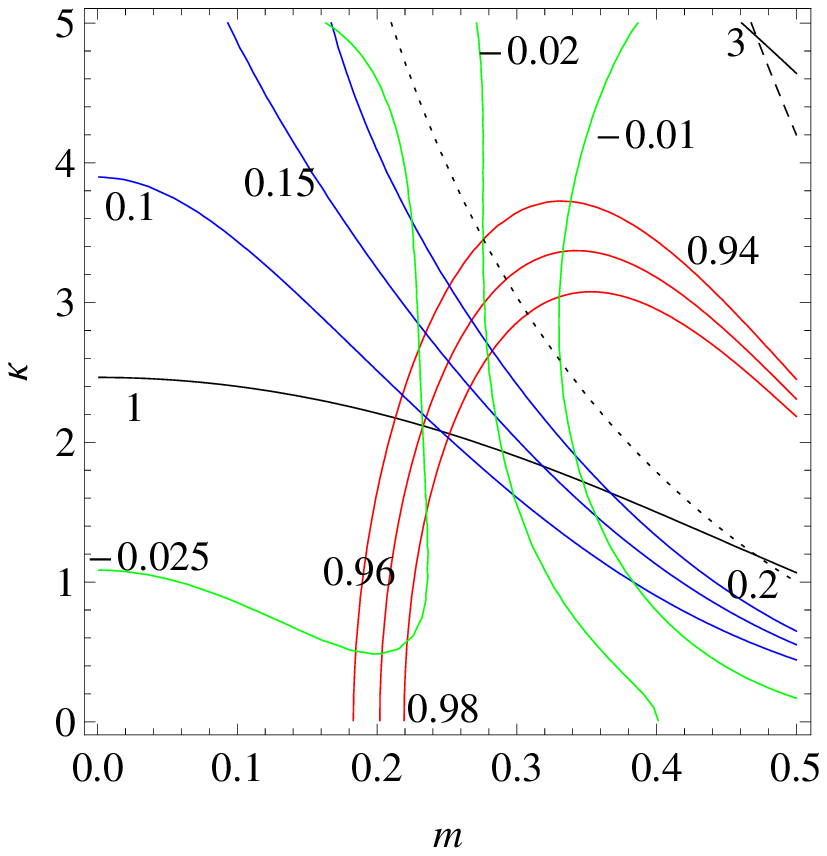, width=9cm,height=9cm,angle=0}
\end{center}
\caption{
Various contours for $N=6$.  }
\label{Fig:6}
\end{figure}

In Figs.~\ref{Fig:50} - \ref{Fig:6}, 
we show the resultant inflationary perturbation indices for various $N$s
 in the $m-\kappa$ plane. 
The blue, red and green lines are contours of  $r_T = 0.1, 0.15, 0.2$, and 
 $n_s = 0.94, 0.96, 0.98$ and
 $\alpha_s = -0.025, -0.02, -0.01$, respectively.
A green line does not appear if the value of $|\alpha_s|$ is very small.
Dashed and dotted lines are used to indicate
 the supergravity correction contribution to the total scalar potential
\begin{equation}
 \frac{\frac{1}{2}m^2\sigma^2}{1+\frac{1}{2}m^2\sigma^2},
\end{equation}
 for $0.5$ and $0.1$, respectively. In the region above the dashed line,
 the false vacuum energy contribution to the total scalar potential is not dominant
 and the model reduces to the chaotic inflation-like.
Gray lines are the contour of the field value of $\sigma$ for each $N$.

Figure~\ref{Fig:50} is for $N=50$, which is large enough to solve
 the horizon and flatness problems only by this inflation.
The region of the $m\rightarrow 0$ and a small $\kappa$ is
 the most well studied part, which predicts the well known results
 $n_s \simeq 0.98$ and negligible $r_T$.
On the other hand,
 although the large $m$ region appears to predict large $r_T$ and $n_s \simeq 0.96$,
 this actually corresponds to the usual quadratic chaotic inflation
 where the false vacuum potential energy is negligible
 and the field value is of ${\cal O}(10)$.

Next, we present the $N$ dependence by showing $N=20$ case in Fig.~\ref{Fig:20}.
By comparing the previous $N=50$ and the next $N=10$ cases,
 one can see the prediction changes as $N$ is varied. 
Since $N$ is reduced, the slow roll parameters increase.
As a result, for a given $\kappa$ and $m$,
 $n_s$ becomes smaller and $r_T$ becomes larger.
However, for $m \gtrsim 0.3$, again
 the false vacuum energy is not dominant.

Now, we consider the case that
 the observed cosmological scale corresponds to $N=10$ in the hybrid inflation,
 shown in Fig.~\ref{Fig:10}.
We see that $n_s \simeq 0.96$ and $r_T = (0.1-0.2)$ are realized
 for $(m, \kappa)\simeq (0.2, (3 - 4) )$, the running is small though.
Here it is clear to see such a large $r_T$ is obtained
 even if the false vacuum energy is dominated.

Finally, in Fig.~\ref{Fig:6},
 we present a case with $N=6$, where
$(n_s, r_T, \alpha_s) \sim (0.96, 0.15, -0.02)$ is obtained
 for $(m, \kappa) \simeq (0.25, 2.5)$.
This offers a resolution of 
 the tension between BICEP2 ($r_T \simeq 0.2$) and Planck ($r_T < 0.11$)
 due to the large enough running
 $\alpha_s$~\footnote{For $\kappa \lesssim {\cal O}(0.1)$
 with a supergravity correction of nonvanishing $m$,
 a large running $\alpha_s$ and wide range of $n_s$ can be obtained. 
This is essentially same manner as in Ref.~\cite{AfterWMAP1-1,AfterWMAP1-2} 
 to reconcile the first indication of a large running by WMAP(2003)~\cite{WMAP1-1,WMAP1-2}.}.

\subsection{Following inflation}

As we have seen in the previous subsection, 
 we can obtain $n_s \simeq 0.96$ and $r_T = {\cal O}(0.1)$,
 if the cosmological scale $k_*$ corresponds to $N \lesssim 10$
 of the hybrid inflation.
Since this short inflation cannot solve the cosmological problems
 in the standard Big Bang cosmology,
 we need additional inflationary expansion with
 the number of e-folds about $50$.
In principle, any inflation can play a role of this.

The double inflation scenario by employing
 the second low scale small field inflation model
 has been investigated many time in literature, for instance,
 referred in Introduction. 
This is one possible scenario. 
For superpotentials for small field inflation,
 see e.g. Refs.~\cite{Izawa:1996dv,Izawa:1998rh}.
During and after hybrid inflation, 
 due to the supersymmetry breaking effects
 by the inflaton of the hybrid inflation and supergravity effects,
 the initial condition for the second small field inflation can 
 be set~\cite{Izawa:1997df}.
After the energy densities of the oscillating $\sigma$ and $\phi$
 decreased and the potential energy for the second inflation dominates,
 the second inflation can take place.
Then, the desired additional number of e-fold and
 an acceptable amplitude of density perturbation can be obtained
 by tuning the potential curvature (see e.g., Refs.~\cite{PBH2-1,PBH2-2,PBH2-3,AfterWMAP1-1,AfterWMAP1-2}).

As mentioned in the Introduction,
 thermal inflation driven by a flaton,
 which would be identified with a flat direction in a supersymmetric model,
 is also a candidate of sequel inflation~\cite{Lyth:1995ka-1,Lyth:1995ka-2,Lyth:1995ka-3}.
In this case, $\sigma$ and $\phi$ reheat the Universe once and
 the thermal effect sets the initial condition for thermal inflation.
After the cosmic temperature drops,
 the potential energy of the flaton induces the thermal inflation.


\section{Summary and discussion}

In this paper, we have examined the possibility
 of viable hybrid inflation generating a large tensor-to-scalar ratio
 as BICEP2 indicated.
The result is that it is possible if an additional inflation follows
 the hybrid inflation whose number of e-fold is not large enough
 to solve the horizon and flatness problems,
 which is crucial and essential assumption. 
Provided the additional inflation successfully works, 
 the observed cosmological scale would correspond to
 a few  $ \lesssim N \lesssim 10$ of the hybrid inflation.
In this case, $n_s \simeq 0.96$ and $r_T = {\cal O}(0.1)$ can
 be realized at the cosmological scale.
One observation here is that if we work on double inflation scenario,
 we do not need super-Planckian field value of the inflaton.
As shown in Figs.~\ref{Fig:10} and \ref{Fig:6},
 the order of Planck field value is sufficient.
Another feature is that a large negative $\alpha_s$ can be
 obtained for $N\sim 6$. 
This would offer a possibility
 to resolve the tension between BICEP2 and Planck.

In double inflation scenario, the present large scale exits the Hubble horizon
 during the high scale hybrid inflation and the small scale does during
 the following low scale inflation.
Hence, a cosmological implication of double inflation
 is strong scale dependence of amplitude of gravitational wave mode.


\section*{Acknowledgments}
This work was supported in part by the Grant-in-Aid for Scientific Research 
No.~25400252 (T.K.) and on Innovative Areas No.~26105514 (O.S.)
 from the Ministry of Education, Culture, Sports, Science and Technology in Japan. 
%



\end{document}